\documentclass[fleqn,usenatbib]{mnras}

\usepackage{newtxtext,newtxmath}

\usepackage[T1]{fontenc}

\DeclareRobustCommand{\VAN}[3]{#2}
\let\VANthebibliography\thebibliography
\def\thebibliography{\DeclareRobustCommand{\VAN}[3]{##3}\VANthebibliography}

\usepackage{graphicx}	
\usepackage{amsmath}	
\usepackage{hyperref}
\usepackage{threeparttable}

\title[Dwarf Galaxies at $z\sim2$]{The Rest-UV Spectral Properties of 
Dwarf Galaxies at $z\sim2$}

\author[C. Snapp-Kolas et al.]{
Christopher Snapp-Kolas,$^{1,{\href{https://orcid.org/0000-0002-9593-0053}{\includegraphics[height=0.3cm]{Images/orcidpic.pdf}}}}$\thanks{E-mail: csnappkolas@gmail.com}
Brian Siana,$^{1,{\href{https://orcid.org/0000-0002-4935-9511}{\includegraphics[height=0.3cm]{Images/orcidpic.pdf}}}}$
Timothy Gburek$^{1,{\href{https://orcid.org/0000-0002-7732-9205}{\includegraphics[height=0.3cm]{Images/orcidpic.pdf}}}}$
Anahita Alavi$^{2,{\href{https://orcid.org/0000-0002-8630-6435}{\includegraphics[height=0.3cm]{Images/orcidpic.pdf}}}}$
Najmeh Emami$^{3,{\href{https://orcid.org/0000-0003-2047-1689}{\includegraphics[height=0.3cm]{Images/orcidpic.pdf}}}}$
\newauthor Johan Richard$^{4,{\href{https://orcid.org/0000-0001-5492-1049}{\includegraphics[height=0.3cm]{Images/orcidpic.pdf}}}}$
Daniel P. Stark$^{5}$
\\
$^{1}$Department of Physics \& Astronomy, University of California, Riverside, CA 92521, US\\
$^{2}$IPAC, California Institute of Technology, 1200 E. California Boulevard, Pasadena, CA 91125, USA\\
$^{3}$Minnesota Institute for Astrophysics, University of Minnesota, Minneapolis, MN, 55455, USA\\
$^{4}$Univ Lyon, Univ Lyon1, Ens de Lyon, CNRS, Centre de Recherche Astrophysique de Lyon UMR5574, F-69230, Saint-Genis-Laval, France\\
$^{5}$Steward Observatory, University of Arizona, 933 N Cherry Ave, Tucson, AZ 85721, USA\\
}

\date{Accepted XXX. Received YYY; in original form ZZZ}

\pubyear{2023}

\begin{document}
\label{firstpage}
\pagerange{\pageref{firstpage}--\pageref{lastpage}}
\maketitle

\begin{abstract}
    Rest-UV spectroscopy can constrain properties of the stellar populations, outflows, covering fractions, and can indirectly constrain the Lyman continuum escape fraction of galaxies. Many works have studied the rest-UV spectra of more massive star forming galaxies and low-mass galaxies selected via strong nebular line emission or via Ly$\alpha$ emission. However, studies of rest-UV spectroscopy have yet to be done on an unbiased sample at low mass during the epoch of peak star formation ($z\sim2$). We present a stacked rest-UV spectrum of a complete sample of 16 dwarf galaxies ($\rm \langle log(M^*/M_\odot)\rangle_{median} = 8.2$) at $z\sim2$. The rest-UV Keck/LRIS spectroscopy is complemented by rest-optical Keck/MOSFIRE spectroscopy and Hubble photometry. We find generally larger Ly$\alpha$ equivalent widths ($\rm EW_{Ly\alpha} = 11.2\;$\AA) when compared with higher mass ($\rm \langle log(M^*/M_\odot)\rangle_{median} = 10.3$) composites from KBSS ($\rm EW_{Ly\alpha} = -5\;$\AA). The average low- and high-ionization absorption line EWs ($\rm EW_{LIS}$ and $\rm EW_{HIS}$, respectively) are weaker ($\rm EW_{LIS}$=-1.18 \AA, $\rm EW_{HIS}=$-0.99 \AA) in dwarf galaxies than in higher mass galaxies ($\rm EW_{LIS}$=-2.04 \AA, $\rm EW_{HIS}=$-1.42 \AA). The LIS absorption lines are optically thick and is thus a good tracer of the neutral hydrogen covering fraction. Both higher $\rm EW_{Ly\alpha}$ and lower $\rm EW_{LIS}$ measurements imply that the escape fraction of ionizing radiation is larger in lower-mass galaxies at $z\sim2$.
\end{abstract}
\section{Introduction}
 The evolution of galaxies over cosmic time is heavily dependent on the cycle of baryons in the intersteller and circumgalactic media (ISM and CGM, respectively). The motion and distribution of gas, dust, and metals in the ISM and CGM dictate the star-formation efficiency and ionizing escape fraction of galaxies. The rest-UV spectra of galaxies can be used to constrain the motion and distribution of gas, dust, and metals. This has been well studied in high-mass ($\rm log(M^*/M_\odot) > 9$) galaxies at high redshift \citep[e.g.][]{Shapley2003,Du2018,Weldon2022}. However, there is yet to be a systematic study of the rest-UV spectra of a complete sample of low-mass galaxies at high redshift.

The rest-UV spectra of galaxies contain complex and informative emission and absorption line profiles. Some of the earliest uses of rest-UV spectroscopy for high redshift galaxies confirmed the redshifts of photometric dropout samples \citep[e.g.][]{Steidel1996a,Lowenthal1997}. More recently, properties of the stellar populations and the gas in the interstellar medium (ISM) have been inferred from the profiles and equivalent widths (EWs) of the emission and absorption lines present in the rest-UV spectra. In particular, the low- and high-ionization absorption lines (LIS and HIS respectively) can trace the neutral and ionized components of the ISM respectively \citep[e.g.][]{Hashimoto2013,Du2018,Du2021,saldana-lopez2022}. Complementary photometry can serve to correlate the measurements of the lines with galaxy physical properties such as UV luminosity and stellar mass \citep[e.g.][]{Shapley2003,Du2018}. 

A primary use of rest-UV spectroscopy is the study of the Ly$\alpha$ emission line. Because Ly$\alpha$ is a resonant line transition in neutral hydrogen, its profile is affected by the existence of even small amounts of neutral hydrogen within a galaxy. However, this also makes Ly$\alpha$ a good tracer of the covering fraction of neutral hydrogen \citep[e.g.][]{Matthee2021,Snapp-Kolas2022}. The escape of Ly$\alpha$ photons is also used as a proxy of the Lyman continuum escape fraction which is relevant for the ionizing background radiation of the intergalactic medium \citep[e.g.][]{Matthee2021a}. Moreover, several studies \citep[][etc.]{Jones2012,Berry2012b,Shibuya2014,Oyarzun2017} have shown an anti-correlation between Ly$\alpha$ EW and LIS EW. Together this suggests greater amounts of Lyman continuum photons escaping from galaxies with higher Ly$\alpha$ EW and lower LIS absorption EW. Because Ly$\alpha$ emitters (LAEs) are typically less massive \citep[e.g.][]{Cullen2020,Pucha2022} this may imply greater Lyman continuum escape fractions for low mass galaxies. However, there is not a study of a complete sample of low-mass galaxies to confirm this trend.

Some ions in the ISM have ionization potentials lower than that of neutral hydrogen and therefore typically exist within regions of neutral hydrogen of sufficient column density to shield these ions from ionizing radiation. Because of this, these lines are often used to trace the outflows of neutral gas in the ISM \citep{Shapley2003,Steidel2010a,Jones2012,Hashimoto2013,Du2018,sugahara2019}. Generally speaking the typical velocity offset of LIS absorption lines remains the same across galaxy properties studied in the literature and is found to be $\sim -180\;\rm km \; s^{-1}$ \citep{Steidel2010a,Hashimoto2013,Shibuya2014}. LIS absorption equivalent width (EW) is also often correlated with Ly$\alpha$ EW \citep{Jones2012,Du2018} such that weaker absorption is associated with stronger Ly$\alpha$ emission. This likely implies greater avenues of escape for Lyman continuum radiation. In fact, \citet{saldana-lopez2022} indirectly show an anti-correlation between LIS absorption EW and the escape fraction of Lyman continuum radiation (see their figures 6 and 9). All of these properties are well studied at higher masses \citep[$\rm > 10^9\;M_\odot$,][]{Shapley2003,Berry2012b,Jones2012,Du2018} or for samples chosen to have large Ly$\alpha$ EWs \citep[e.g.][]{Hashimoto2013,Shibuya2014}, but little work has been done on a complete sample at low-mass to see if these trends hold for dwarf galaxies. 

Outside regions of dense neutral hydrogen the elements can be subject to harder radiation and more highly ionized ions can be produced. These ions have ionization potentials well above that of neutral hydrogen and therefore will trace ionized regions of the ISM. \citet{Shapley2003} show that the HIS absorption line EW is constant in all of their non-LAE bins, but in their LAE bin the depth of the HIS EW decreases. \citet{Du2018} show similar behavior for their $z\sim2$ sample with the highest Ly$\alpha$ EW bin having weaker HIS absorption, arguably due to greater numbers of Ly$\alpha$ emitters contributing to the stack in that bin. Both argue that there is an insignificant change in the HIS EW strength with Ly$\alpha$ EW, but each display a sudden change in the strength of the HIS EW for LAEs. Since LAEs are less massive than Lyman break galaxies (LBGs) generally \citep[e.g.][]{Cullen2020,Pucha2022} it may be the case that lower mass galaxies have lower HIS EWs.  

In this work we utilize photometric and spectroscopic data to study the UV spectral properties of dwarf galaxies and compare with the better-studied more massive galaxies. The remainder of the paper is organized as follows. In $\S$\ref{Observations}, we briefly review the observations and describe additional spectral energy distribution (SED) fitting, sub-sample selection, and describe the spectral stacking methodology. In $\S$\ref{Results}, we present measurements of the Ly$\alpha$, LIS, and HIS EWs and velocity offsets. In $\S$\ref{discussion}, we compare with measurements found in the literature and discuss the implications of these comparisons. In $\S$\ref{Summary}, we summarize our findings. We adopt a $\Lambda$CDM cosmology with $\Omega_m$ = 0.3, $\Omega_{\Lambda} = 0.7$, and $h = 0.7$ throughout the paper and all magnitudes are in the AB system \citep{Oke1983a}. In this work all EWs are given in the rest frame. Emission lines are taken to have positive EWs and absorption lines are taken to have negative EWs. We use the convention of positive velocities indicating a redshift and negative velocities indicating a blueshift.

\section{Observations, Data Reduction, and Sample Selection}
\label{Observations}
\subsection{Observations \& Data Reduction}
This work is based on data discussed in \citet{Snapp-Kolas2022} and the details of the observations and reduction can be found therein. We briefly review the basics of the sample here. Our sample consists of a subset of dwarf galaxies observed with deep HST photometry and low resolution rest-UV spectroscopy with Keck/LRIS \citep{Oke1995}. All of our galaxies are behind three lensing clusters (Abell 1689, MACS J1149, and MACS J0717) in order to observe the faintest UV continua possible with LRIS. Exposure times varied from 4500s-13000s in our 11 masks according to the conditions during observations. The typical seeing of our data is $\sim 1''$. Objects were selected to have visual magnitudes brighter than m$_{\rm F625W}<26.3$ and photometric redshifts in the range $1.5<z<3.5$. Photometric redshifts are determined from our HST photometry \citep[see][for details]{Alavi2014,Alavi2016}. Galaxies with high magnification were given priority when creating slit masks. The LRIS data were reduced and extracted using a modified version of the PypeIt v1.x reduction pipeline \citep{pypeit:joss_arXiv}, which performs flat-fielding, wavelength calibration, cosmic ray rejection, sky subtraction, and optimal extraction of 1D spectra. We correct these extracted spectra for slit-losses using Hubble photometry. The spectral coverage is 3100 \AA-5600 \AA\ and the spectral resolution is $\rm R\sim500$. After removing spectra with contamination from internal reflections we have a parent sample of 127 spectra. After combining multiple images we have a parent galaxy sample of 89 galaxies. Keck/MOSFIRE \citep{McLean2010,McLean2012} rest-optical spectra were also obtained for a portion of these galaxies, which we use to confirm the redshifts and measure the H$\alpha$ emission line of the galaxies used in this work. The details of the MOSFIRE spectra can be found in \citet{gburek2023}, but we shall summarize the spectroscopy here as well. The spectra were observed using a $2.5''$ ABBA dither pattern in the Y-,J-,H-, and K-bands with targets selected to have typical strong nebular emission lines (i.e. [OII] $\lambda\lambda$3726, 3729, H$\beta$, [OIII] $\lambda\lambda$4959, 5007, H$\alpha$, and [NII] $\lambda\lambda$6548, 6583) in 9 masks. The average exposure times for the Y-, J-, H-, and K-bands were 96 min., 81 min., 85 min., and 82 min. with resolutions of $\rm R = $ 3388, 3318, 3660, and 3610 respectively\footnote{\url{https://www2.keck.hawaii.edu/inst/mosﬁre/grating.html}}. The spectra were reduced using the MOSFIRE Data Reduction Pipeline\footnote{\url{https://keck-datareductionpipelines.github.io/MosﬁreDRP/}} (DRP). 1D spectra were then extracted from the reduced 2D spectra using \texttt{BMEP}\footnote{\url{https://github.com/billfreeman44/bmep}} from \citet{Freeman2019}.

\subsection{SED fit}
We fit SEDs to our galaxies within the IR footprint of our Hubble photometry  in \citet{Alavi2014,Alavi2016}. In that work we make use of the code \texttt{FAST}\footnote{\url{https://w.astro.berkeley.edu/~mariska/FAST.html}} \citep{Kriek2009} and we fit \citet{Bruzual2003MNRAS.344.1000B} stellar population synthesis (SPS) models assuming constant star formation histories (SFHs), a \citet{chabrier2003} initial mass function (IMF), stellar metallicities of either $\rm 0.2 Z_{\odot}$ or $\rm 0.4 Z_{\odot}$, and an SMC dust attentuation curve. Three galaxies in this sample for this work did not have SED fits, as they were not observed in all of the filters of the \citet{Alavi2014,Alavi2016} sample. Nevertheless, the photometry was sufficient for performing SED fits. For these galaxies we made use of \texttt{BAGPIPES}\footnote{\url{https://bagpipes.readthedocs.io/en/latest/index.html}} \citep{Carnall_2018} to perform SED fitting on these galaxies. The same assumptions were used as in \texttt{FAST}, but a \citet{Kroupa2004} IMF is used instead as this is fixed in the \texttt{BAGPIPES} code. We note that the \citet{chabrier2003} and \citet{Kroupa2004} IMF's differ primarily at the low mass end ($\rm log(M^*/M_\odot) < 1$), and differences in the two are small in terms of the number of stars of a given mass. Therefore, we argue that the two assumptions will have little difference in the estimated mass which is the primary use of the SED fits in this work.

\subsection{Sub-sample Selection}

For the purposes of this study we aim to choose a sample that is complete in terms of its star-forming properties. We place the following constraints on our sample to accomplish this goal \citep{Snapp-Kolas2022}:
\begin{itemize}
    \item Only galaxies with confirmed redshift from our rest-optical spectra are kept. These redshifts are determined from H$\alpha$ and [OIII]$\lambda\lambda5007$. 
    \item We require that H$\alpha$ be observable within the MOSFIRE H, J, or K bands. This limits our observations to galaxies with $z \lesssim 2.6$.
    \item We remove galaxies that show blending with other nearby galaxies in the HST photometry
    \item We remove galaxies with magnification $\mu > 30$ behind Abell 1689 and $\mu>15$ for MACS J0717 and MACS J1149. This is to ensure the galaxies in the sample are sufficiently far away from the critical line, to avoid differential magnification across a galaxy.
    \item We remove galaxies with large slit losses (defined empirically to be a slit loss of 55\% or more in the LRIS and MOSFIRE spectra). Slit losses larger than this reduce confidence that the spectrum is representative of the whole galaxy, as a majority of the galaxy light lies outside of the slit.
    \item We additionally remove any galaxies that could not have been detected at our 3$\sigma$ H$\alpha$ sensitivity limit at $\rm Log(H\alpha/UV) = 13.4$ given the magnification of each galaxy \citep[see][]{Snapp-Kolas2022}, and could not have been detected in H$\alpha$ below the lower end of the extrapolated star-forming main sequence of \citet{Sanders2021} given the sensitivity of our Keck/MOSFIRE spectra. This is done to remove a bias towards galaxies that have large H$\alpha$ luminosities. 

\end{itemize}
\begin{figure}
    \centering
    \includegraphics[height=7cm]{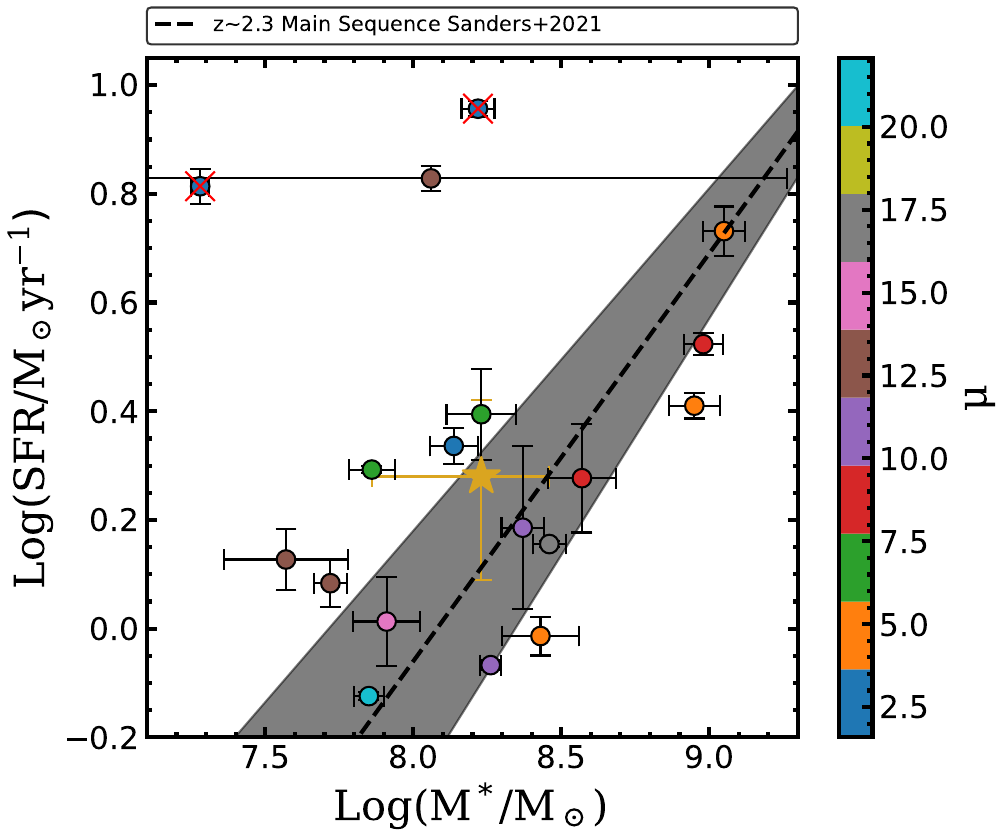}
    \caption{Galaxy main sequence for our dwarf galaxy sample. Points are color coded according to the magnification of each galaxy (see color bar). The main sequence extrapolated from the empirical \citet{Sanders2021} relation (eq. (3) in their paper) is plotted for reference along with the 1$\sigma$ errors shaded in grey. The errors on the individual measurements are derived from the SED fitting codes. The median SFR and mass are plotted together with a golden star. The error bars on the star represent the 25th and 75th percentiles of the sample. Galaxies excluded because they would have H$\alpha$ luminosites below our H$\alpha$ sensitivity limit (adjusted by redshift and magnification) are marked with red x's.}
    \label{fig:main_seq}
\end{figure}

Figure \ref{fig:main_seq} shows the star-formation rate (SFR) vs. stellar mass of the remaining sample after the above considerations. The galaxies that have red x's could NOT have been detected in H$\alpha$ below the main sequence and are therefore removed from the sample. Points are color-coded according to their magnification and the gold star represents the median SFR and Mass of the sample. The error bars on the star represent the 25th and 75th quintiles of the sample for each variable. Error bars on the individual galaxies are derived from the SED fits. These considerations leave us with a final complete sample of 16 dwarf galaxies. The sample is consistent with lying on the main sequence within 1$\sigma$ of the \citet{Sanders2021} relation.

\section{Results}
\label{Results}
Thanks to lensing we are able to probe down to $\rm M_{UV}\lesssim-17$ at $z\sim2$, which is two magnitudes fainter than the KBSS sample ($\rm M_{UV}\lesssim-19$) of \citet{Du2018}. However, given the low luminosities of our sample, the S/N in the continuum is still too low to detect the LIS and HIS absorption features in individual spectra. Therefore, we perform a median stack of all the galaxies in our final sample to produce a typical dwarf galaxy spectrum to compare with larger mass galaxies. We normalize the individual spectra with a power law of the form $f_\lambda \sim \rm \lambda^\beta$ before stacking, so that the continuum has a constant value of unity. Because the sample consists of galaxies at various redshifts between 1.6 and 2.6 the spectral coverage of the stack is limited to 1200 \AA\ to 1560 \AA to ensure that all galaxies contribute to the stack at all wavelengths.

\begin{figure*}
    \centering
    \includegraphics[width=17cm]{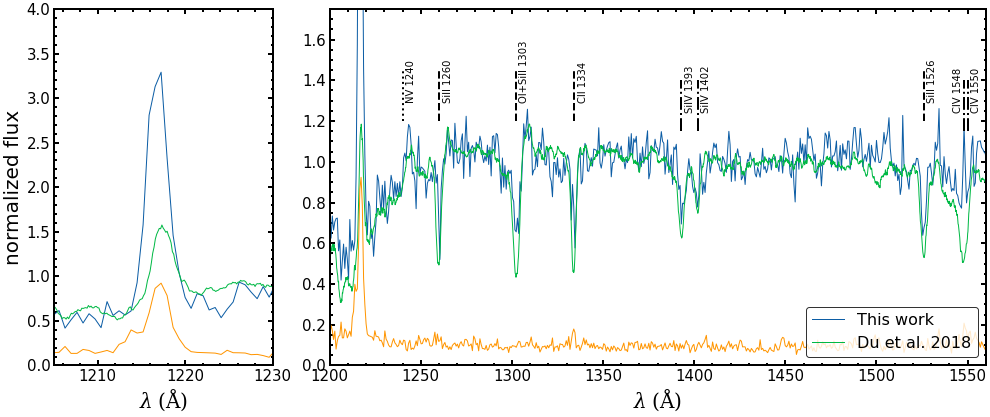}
    \caption{The stacked spectrum of the 16 dwarf galaxies in our sample is plotted in blue. The $z\sim 2$ stack of \citet{Du2018} is plotted in green. To compare the depths of the absorption lines we normalize the spectrum of \citet{Du2018} by the continuum, which we estimate using a running average with a width of 100 \AA. The error spectrum is plotted in orange and relevant lines are labeled in the figure. Low-ionization absorption lines are labeled with dashes, high ionization lines are labeled with dot-dashes, and the NV line is labeled with a dotted line. Left: A comparison of the Ly$\alpha$ emission line. Right: the rest-UV continuum and absorption lines of the same stacks. Generally, the Ly$\alpha$ emission line is stronger, and both the LIS and HIS absorption lines are weaker than the \citet{Du2018} stack.} 
    \label{fig:stack}
\end{figure*}

    Each spectrum in the stack contains statistical uncertainty given by the error spectrum. In addition, galaxies have a wide range of properties and given the size of our sample there is likely inherent variance in the stack do to this variance in galaxy properties. To account for both the statistical uncertainty and the variance in galaxy properties we perturb each spectrum according to its error spectrum and randomly select with replacement from our sample. We then median stack these mock spectra. We repeat this process 1000 times and then take the standard deviation of the flux at each wavelength bin to be the error spectrum of our median stack. We use this error spectrum to produce the errors on our measurements of the absorption lines and the Ly$\alpha$ emission line. The normalized stack is shown in figure \ref{fig:stack} along with the normalized $\rm z\sim 2$ stack of \citet{Du2018}.

     Our stacked spectrum spans the rest-frame wavelength range of 1200 \AA-1560 \AA\ and therefore probes the absorption lines of low and high ions and common emission lines such as the Si II fine structure emission lines, CIV emission, and N V $\rm\lambda$1240. We also are able to measure the Ly$\alpha$ emission line EW and velocity offset relative to systemic to compare with \citet{Du2018}. 

We measure the LIS absorption lines by performing a parametric fit of the Si II $\rm \lambda$1260, O I $\rm\lambda$1302+Si II $\rm\lambda$1304, C II $\rm\lambda$1334, and Si II $\rm\lambda$1527 lines. We assume Gaussian profiles and require the offset from systemic and the width of the lines to be the same for each LIS absorption line. This naturally presumes that all of the low ions occupy the same region of the galaxy. We use this model in a Markov Chain Monte Carlo procedure with PyMC3 to fit the lines to the data. This method takes into account the error spectrum on the medium stack described in section \ref{Observations} and our measurement errors are taken from this fitting procedure. The HIS absorption lines are fit using a similar method but centering the Gaussian profiles on Si IV $\rm\lambda\lambda$1393, 1402 and C IV $\rm\lambda\lambda$1548, 1550 and not requiring the two to be at the same velocity shift or have the same width as in \citet{Shapley2003} and \citet{Du2018}. The SiIV doublet is spectrally resolved, but the CIV doublet includes a P-Cygni profile that originates from stellar winds in massive stars. This makes it difficult to disentangle the stellar and interstellar components of the CIV profile \citep{Du2018}. Additionally, \citet{Rudie2019} observe a different profile for CIV in their $\rm L^*$ galaxies, suggesting it traces an additional phase of the gas. Because of this, the CIV doublet may not be probing the same region of the galaxy as the SiIV doublet. Therefore, we take the HIS absorption line EW to be the average of the SiIV doublet.

We fit the Ly$\alpha$ EW according to the methods of \citet{Du2018}, which we briefly review here. We take the continuum to be the midpoint flux of a line passing through the flux at 1208 \AA\ and 1240 \AA . We then integrate from 1208 \AA\ to 1240 \AA\ to calculate the EW \citep[see][for more details]{Snapp-Kolas2022}. To get the velocity offset of the Ly$\alpha$ profile we parameterize according to a Gaussian on top of a linear continuum. The same MCMC method used for the LIS absorption lines is used for this fit. Our measured EWs and velocity offsets from systemic are listed in table \ref{tab:example_table}. Given our limited resolution in our composite spectrum we are unable to properly characterize the maximum velocity of our galaxies \citep{Vasan2022}. We choose instead to fit a single Gaussian profile to each of our lines and use the line center of the fit to represent the outflow velocities of our galaxies do to our low-resolution spectroscopy ($R\sim 500$). We also measure $\rm EW_{LIS}$ which we define as the mean EW of the Si II $\rm \lambda$1260, O I $\rm\lambda$1302+Si II $\rm\lambda$1304, C II $\rm\lambda$1334, and Si II $\rm\lambda$1527 absorption lines. We define it this way in order to compare with the literature which typically defines $\rm EW_{LIS}$ as such \citep[e.g.][]{Berry2012b,Jones2012,Du2018,saldana-lopez2022}. We measure $\rm EW_{LIS}$ to be -1.18 $\rm \pm 0.14$ \AA\ and find the HIS EW to be -0.99 $\pm 0.24$ \AA for the total stack. We also find that the Ly$\alpha$ EW is 11.2 $\pm 1.1$\AA\ and the Ly$\alpha$ velocity offset from systemic is 302 $\pm 46$ km $\rm s^{-1}$. 
\begin{table}[h]
	\caption{The emission and absorption line EWs and velocity offsets from systemic measured from the total median stack of sixteen dwarf galaxies. We take the convention of positive velocity indicating a redshift and negative velocities indicating a blueshift.}
    \begin{threeparttable}
	\label{tab:example_table}
	\begin{tabular}{lcccc} 
		\hline
		line & EW & $\rm EW_{err}$ & v & $\rm v_{err}$\\
		(Name) & (\AA) & (\AA) & ($\rm km\; s^{-1}$) & ($\rm km\; s^{-1}$) \\
		\hline
        Ly$\alpha$ & 11.2 & 1.1 & 302 & 46  \\ 
        SiII 1260 & -1.03 & 0.26 & -125 & 40 \\ 
        OI 1302 + SiII 1304 & -1.54 & 0.35 & -121 & 39\\
        CII 1334 & -1.12 & 0.30 & -118 & 38 \\
        SiII 1526 & -1.03 & 0.22 & -103 & 33 \\
        SiIV 1393 & -1.20 & 0.37 & -71 & 67 \\
        SiIV 1402 & -0.77 & 0.29 & -71 & 66  \\
        CIV 1548\tnote{1} & -0.76 & 0.31 & -449 & 60 \\
        CIV 1550\tnote{1} & -0.28 & 0.24 & -448 & 60\\
		\hline
	\end{tabular}
    \begin{tablenotes}
       \item [1] It is unclear whether the absorption features in the stack are in fact interstellar CIV absorption. It is possible that there are large amounts of "filling in" at low velocities from emission from the P-Cygni stellar wind profiles.
    \end{tablenotes}
    \end{threeparttable}
\end{table}
\section{Discussion}
\label{discussion}
To study the mass dependence of properties of dwarf galaxies at a given redshift we compare primarily with the KBSS sample of \citet{Du2018}. 
\subsection{Lyman Alpha}
\begin{figure}
    \centering
    \includegraphics[height=7cm]{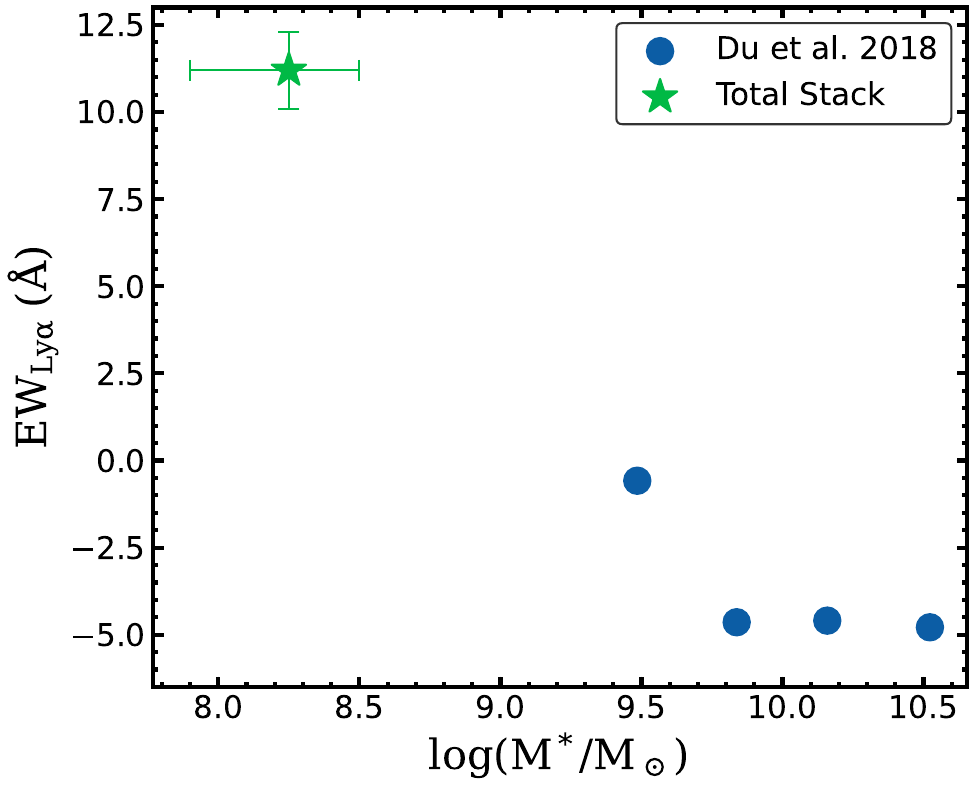}
    \caption{$\rm EW_{Ly\alpha}$ vs. $\rm log(M^*/M_{\odot})$ from stacked spectra at $z\sim2$. Our datum is shown as a green star and the \citet{Du2018} data are shown in blue circles. The sample of \citet{Du2018} is binned according to mass and spectra are stacked in each mass bin. The errors on the KBSS sample are too small to be seen in the figure. For our dwarf galaxy sample the error bars on the Ly$\alpha$ EW are derived from the MCMC fit to the Ly$\alpha$ line as described in section \ref{Results}. The error bars on the mass denote the 25th and 75th percentiles of the distribution respectively. The measured mass value is the median of the total sample. There is a clear increase in $\rm EW_{Ly\alpha}$ with decreasing mass.}
    \label{fig:Lyamass}
\end{figure}

 Figure \ref{fig:Lyamass} shows the Ly$\alpha$ EW as a function of stellar mass. \citet{Du2018} demonstrate a relatively flat trend in $\rm EW_{Ly\alpha}$ with mass down to $\rm log(M^* /M_\odot) \approx 10$. Below this mass they show an increase to $\rm EW_{Ly\alpha}\approx 0$\AA\ at $\rm log(M^* /M_\odot) \approx 9.5$. Our stack continues this trend demonstrating that Ly$\alpha$ will be seen in emission, on average, below $\rm log(M^* /M_\odot) \approx 9$.  We see a clear increase in the Ly$\alpha$ EW with decreasing mass. \citet{Du2018} show that for a given stellar mass $\rm EW_{Ly\alpha}$ will increase with redshift. This is particularly stark at the lowest masses and may suggest that galaxies at the masses of our sample are Ly$\alpha$ emitters (LAEs, $\rm EW_{Ly\alpha}>20$\AA) on average at $z\ge4$. This suggests that lower mass galaxies allow greater amounts of Ly$\alpha$ photons to escape, and therefore suggests greater amounts of ionizing radiation are escaping. This trend towards higher Ly$\alpha$ EWs at lower mass is also observed in other high-redshift samples in the literature \citep[see for e.g.][$3<z<4.6$]{Oyarzun2017}.

\begin{figure}
    \centering
    \includegraphics[height=7cm]{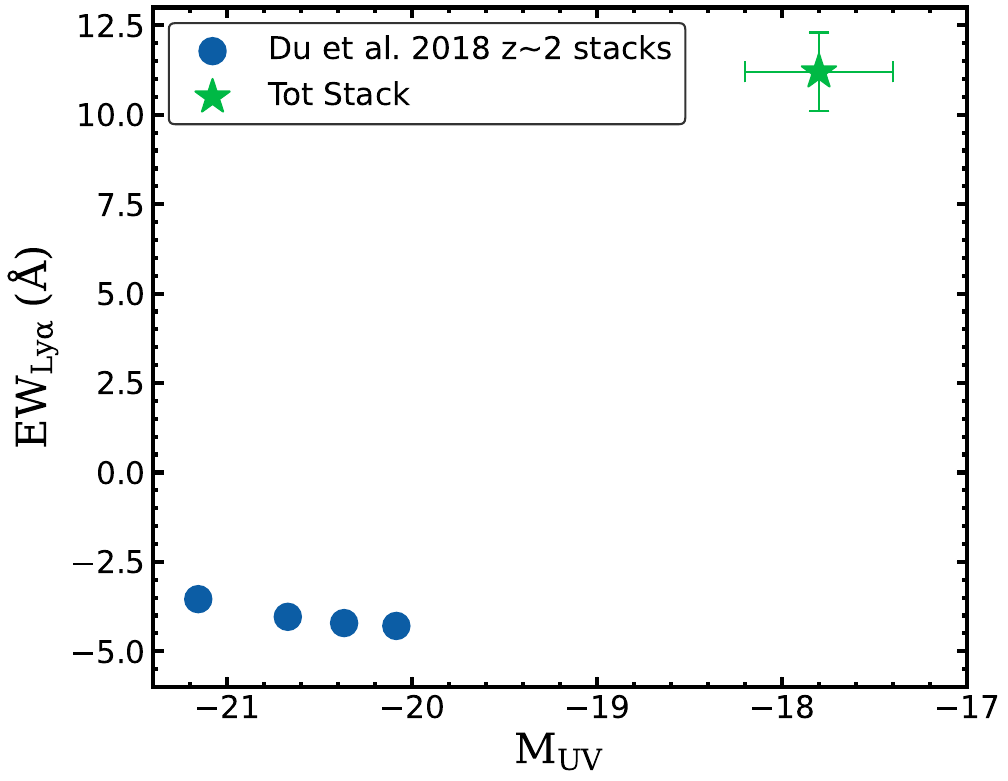}
    \caption{The $\rm EW_{Ly\alpha}$ vs. $\rm M_{UV}$ from stacked spectra at $z\sim2$. The green star is the measurement from our stack and the blue dots are from \citet{Du2018}. Here the \citet{Du2018} sample is binned according to absolute UV magnitude. Again the error bars on the \citet{Du2018} sample are too small to be perceived in the figure and our error bars are calculated in the same manner as in figure \ref{fig:Lyamass}. The error bars on the absolute UV magnitude are the 25th and 75th percentiles of the sample. There is a clear increase in the $\rm EW_{Ly\alpha}$ at fainter UV luminosities.}
    \label{fig:LyaMUV}
\end{figure}

We also compare samples to determine trends in  Ly$\alpha$ EW with absolute UV magnitude. Figure \ref{fig:LyaMUV} is color coded in the same manner as figure \ref{fig:Lyamass}. According to \citet{Du2018} there is a flat trend for $z\sim2$ galaxies in $\rm EW_{Ly\alpha}$ with $\rm M_{UV}$ and each bin shows that Ly$\alpha$ is observed with net absorption ($\rm EW_{Ly\alpha} \sim (-3.5)- (-4.5)$). In our dwarf galaxy sample we find a positive Ly$\alpha$ EW ($\sim 11$ \AA), showing that the typical dwarf galaxy will have net Ly$\alpha$ emission. This is consistent with our earlier work as well \citep{Snapp-Kolas2022}. However, at higher redshift in the \citet{Du2018} sample there is a noted increase in Ly$\alpha$ EW towards fainter absolute UV magnitudes. With our sample added to the analysis of \citet{Du2018} we can conclude that at redshifts $z\sim2-4$ fainter galaxies produce higher Ly$\alpha$ EWs, and generally there exists an absolute UV magnitude at which the typical galaxy shifts from being a net absorber of Ly$\alpha$ photons, to being a net emitter of Ly$\alpha$ photons.

\begin{figure*}
    \centering
    \includegraphics[height=13cm]{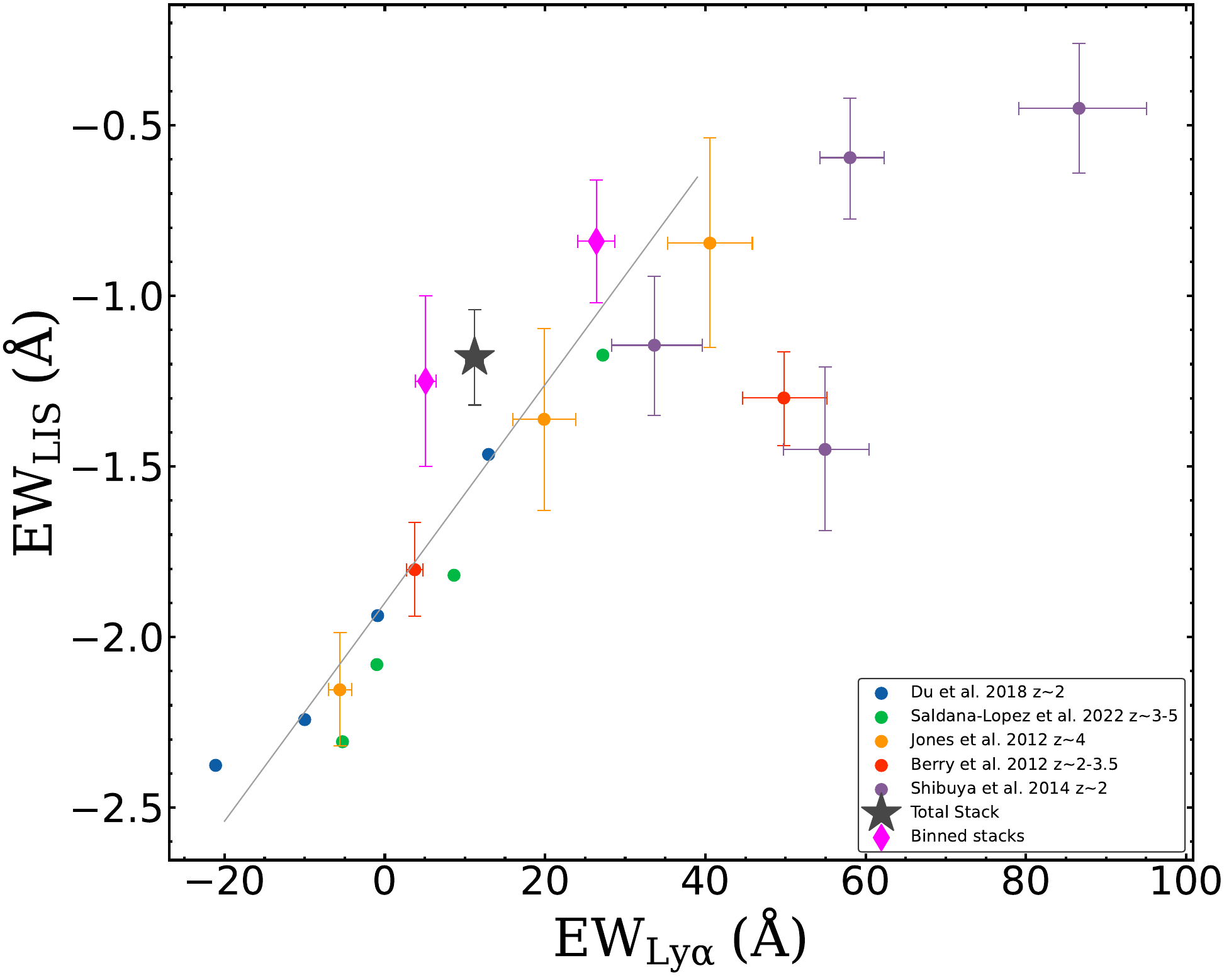}
    \caption{The $\rm EW_{LIS}$ vs. $\rm EW_{Ly\alpha}$ of high-redshift star-forming galaxies. The black star is measured from our total stack with the error on $\rm EW_{LIS}$ determined from the MCMC fitting algorithm (with error spectrum taken into account) and the error on $\rm EW_{Ly\alpha}$ is determined by propagation of error from the error spectrum. The pink diamonds are stacks on 8 galaxies each from our sample of 16 galaxies in two bins of $\rm EW_{Ly\alpha}$ split at 6.2 \AA. The errors are determined in the same manner as the total stack. We also plot values from \citet{Du2018}, \citet{Jones2012}, \citet{Berry2012b}, \citet{Shibuya2014}, and \citet{saldana-lopez2022} in blue, orange, red, purple, and green respectively for comparison with our sample. Each sample bins on $\rm EW_{Ly\alpha}$, and measures values from stacks within the bin, except for \citet{saldana-lopez2022} whose values come from individual LAEs. All of the literature values show a correlation between $\rm EW_{LIS}$ and $\rm EW_{Ly\alpha}$ for $\rm EW_{Ly\alpha} < 45$ \AA . The measurements from our stacks lie above the trends from the literature.} 
    \label{fig:LISLya}
\end{figure*}

 Patches of neutral hydrogen will greatly attenuate the Ly$\alpha$ line strength. Given low-ionization ions exist within regions of neutral hydrogen we expect the strength of Ly$\alpha$ to be correlated with the strength of the LIS absorption lines. Figure \ref{fig:LISLya} shows the LIS EW as a function of the Ly$\alpha$ EW. At the redshift of our sample, \citet[blue]{Du2018} show a clear correlation between the LIS absorption line EW and the Ly$\alpha$ EW. They offer that this supports a physical model of patchy optically thick clumps surrounding star-forming regions. In this scenario the trend between the LIS EW and the Ly$\alpha$ EW is an emergent property of the radiative transfer of Ly$\alpha$ photons. \citet{Du2018} demonstrate that this correlation is invariant up to $z\sim4$. This trend is confirmed by many others in the literature over various galaxy properties and across redshift as shown in figure \ref{fig:LISLya}.

We find smaller absolute values of LIS EW at a given Ly$\alpha$ EW for our sample than those of \citet{Du2018}, \citet{Berry2012b}, and \citet{Shibuya2014}. This suggests that the trend between LIS EW and Ly$\alpha$ EW found in \citet{Du2018} is dependent on stellar mass or UV luminosity as argued by \citet{Jones2012}.  

However, we have stacked on the entire sample for this comparison. In order to more closely compare with the literature we stack two sub-samples chosen on Ly$\alpha$ rest-EW as is done in these other works. We split the sample by the median Ly$\alpha$ EW (6.2 \AA ). The masses of the two subsamples differ by a small amount. The lower and higher Ly$\alpha$ EW samples have masses of $\rm Log(M^*/M_\odot) = 8.35^{+0.32}_{-0.23}$ and $\rm Log(M^*/M_\odot) = 8.07^{+0.32}_{-0.21}$ where the errors denote the 25th and 75th percentiles of the distributions. Each sample has 8 galaxies in the bin, they are stacked in the same manner as the full sample, and the error spectrum is found in the same manner as well. All measurements are performed the same as for the total stack. The results are plotted as pink diamonds in figure \ref{fig:LISLya}. Our data lie above the trend in the literature, suggesting the relationship between $\rm EW_{Ly\alpha}$ and $\rm EW_{LIS}$ may be mass dependent. However, \citet{Du2021} investigate the scatter in this trend using individual galaxy measurements to determine possible drivers of the scatter in this and other correlations. They find that for a fixed Ly$\alpha$ EW the LIS EW will vary based on the metallicity of the galaxy, with weaker LIS absorption occurring in lower metallicity galaxies. As such, the offset of our data from the trend in the literature may be indicative of dwarf galaxies having lower metallicities \citep{gburek2023}. 

\subsection{LIS Absorption Lines and Kinematics}
\begin{figure}
    \centering
    \includegraphics[height=7cm]{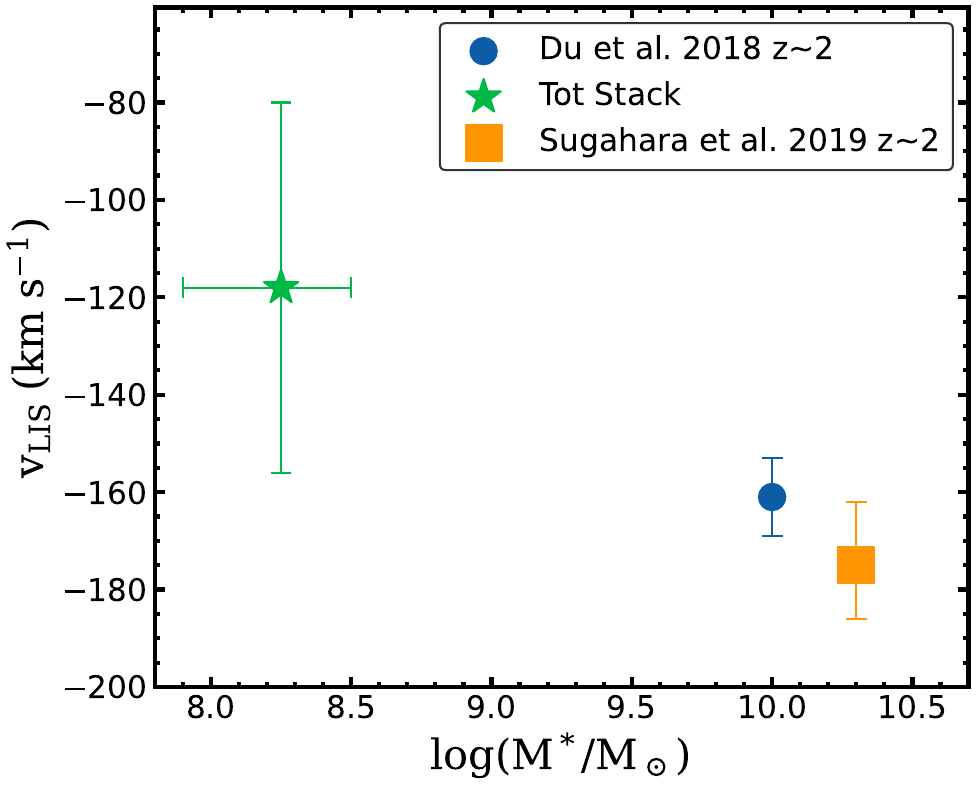}
    \caption{The LIS absorption velocity offset as a function of stellar mass. The green star is from our stack and the blue dot and orange square are from \citet{Du2018} and \citet{sugahara2019} respectively. The errors on the mass for our stack are the same as in figure \ref{fig:Lyamass} and the errors on the velocity are derived from the MCMC algorithm that fit the lines to the spectrum. The measured LIS velocity is lower than \citet{Du2018} and \citet{sugahara2019}, though the uncertainty is comparatively large. The data point from \citet{sugahara2019} is a correction of a measurement first made in \citet{Sugahara2017}}.
    \label{fig:vLISmass}
\end{figure}

 We wish to understand the distribution of neutral hydrogen to better constrain the escape of ionizing radiation. Here we compare our results with the KBSS sample of \citet{Du2018} to find any trends in the LIS absorption lines with mass. The kinematics of ions/gas are often characterized either by the line center velocity of the absorption line or some choice of maximum velocity offset in the absorption line profile. The maximum velocity offset is often chosen to be at 90\% flux relative to the continuum value \citep[e.g][]{Sugahara2017,Weldon2022}.  However, given the resolution of our spectra we choose to use the line center as described in section \ref{Results}. Figure \ref{fig:vLISmass} shows the LIS absorption velocity offset from systemic as a function of mass. To compare with the literature we take the CII 1334 velocity offset to be representative of the LIS absorption lines generally. There is a trend towards lower LIS absorption velocity offsets at lower masses relative to higher mass samples \citep{Du2018,sugahara2019}, though the uncertainties on our measurement render this result tenuous. The  models from the FIRE simulations \citet[][]{Muratov2015} predict that the outflow velocities will be lower in lower mass galaxies. Our results are qualitatively consistent with this model.

\begin{figure}
    \centering
    \includegraphics[height=7cm]{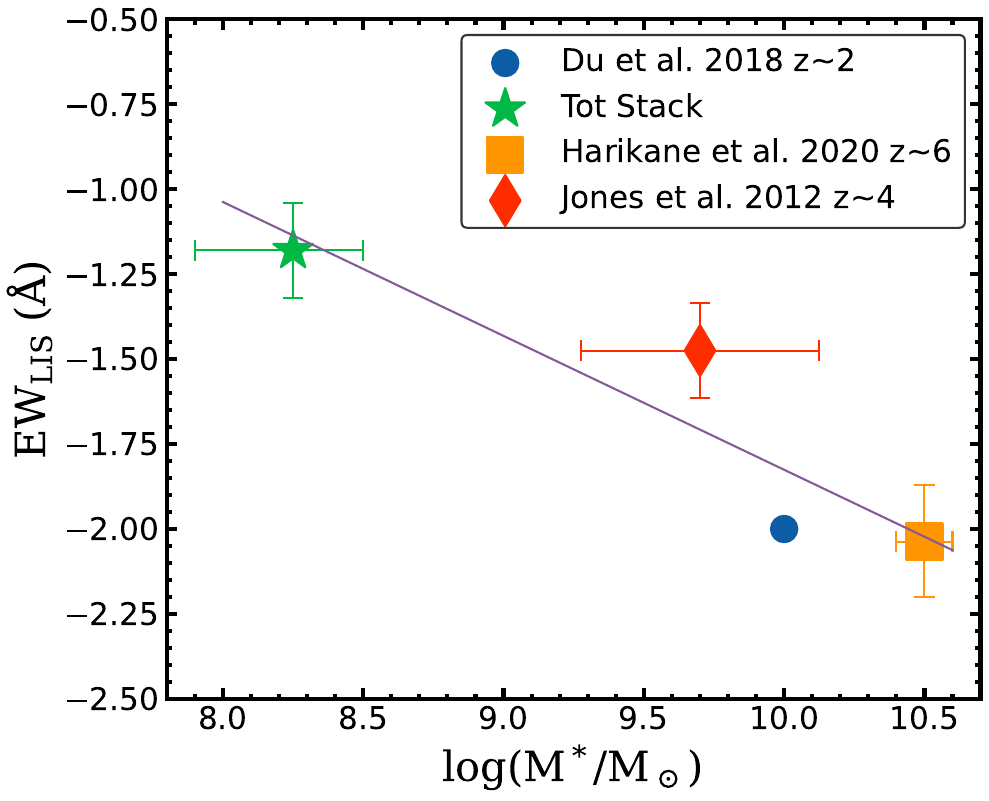}
    \caption{The $\rm EW_{LIS}$ as a function of stellar mass. The green star is our datum, the red diamond is from \citet{Jones2012}, the blue dot is from \citet{Du2018}, and the orange square is from \citet{Harikane2020}. The errors on the mass are as in figure \ref{fig:Lyamass} and the errors on $\rm EW_{LIS}$ are derived from the MCMC fitting algorithm. We also plot a least-squares linear fit to the data points in purple showing that there exists an anti-correlation between $\rm EW_{LIS}$ and $\rm log(M^*/M_\odot)$.}
    \label{fig:LISmass}
\end{figure}

    Figure \ref{fig:LISmass} shows the LIS EW as a function of mass. \citet{Du2018} measure a single value of the LIS EW relative to mass ($\rm EW_{LIS} \approx -2.0$\AA). According to figure \ref{fig:LISLya} \citep[see also][figure 5]{Du2018} there is little to no dependence on the LIS absorption line EW with redshift. Therefore, there should be little variance in any relation with mass due to redshift. With this in mind we also compare with \citet{Jones2012} and \citet{Harikane2020} and observe an apparent anti-correlation. Our datum shows weaker LIS EW at lower mass and further indicates an anti-correlation between LIS EW and mass. Given this we fit an empirical relation between the LIS absorption EW and the log stellar mass of the form:
    \begin{equation}
        {\rm EW_{LIS}} = a_1{\rm log(M^*/M_\odot)} + a_2
    \end{equation}
    We use numpy's polyfit algorithm to perform a least squares fit of this relation to the observed data. We find $ a_1 = -0.39\pm0.12$ and $a_2 = 2.12\pm1.20$. The best fit line is plotted in figure \ref{fig:LISmass}. Additionally, we observe the ratio of silicon lines to be $\rm EW_{SiII1260}/EW_{SiII1526} = 1.0\pm0.3$, which is consistent with the LIS gas being optically thick and the absorption profiles being saturated. Seeing that saturated LIS absorption lines trace neutral hydrogen covering fractions we conclude that lower mass galaxies have lower covering fractions.

\subsection{HIS absorption lines}
\begin{figure}
    \centering
    \includegraphics[height=7cm]{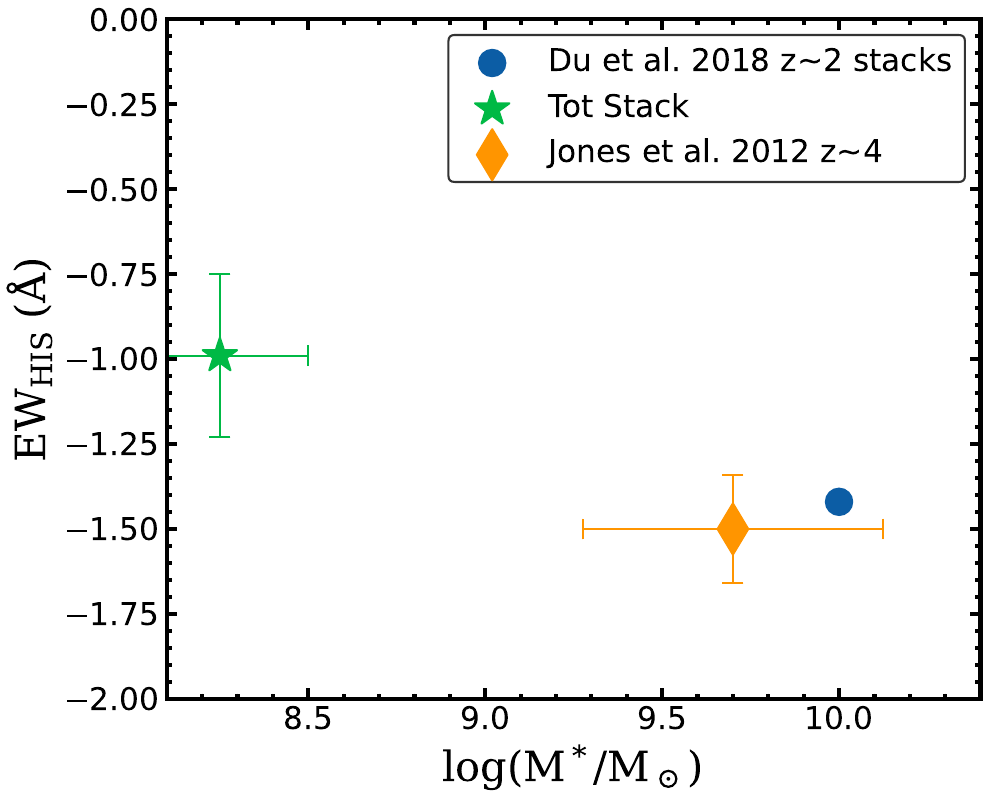}
    \caption{$\rm EW_{HIS}$ vs. stellar mass. The green star is our datum, the blue dot is from \citet{Du2018}, and the orange diamond is from \citet{Jones2012}. The errors on our datum are determined in the same manner as figure \ref{fig:LISmass}. At $z\sim2$ there is less HIS absorption at lower mass. There is no indication of a change in the depth of the HIS absorption lines with redshift for high mass galaxies. Therefore, the depth of HIS absorption lines appear to depend primarily on mass and do not evolve with time.}
    \label{fig:HISmass}
\end{figure}

The relative abundance of high ions and low ions is indicative of the ionization state of the ISM and of outflowing gas. A more highly ionized ISM is likely to accommodate the escape of ionizing radiation. In figure \ref{fig:HISmass} we plot the HIS absorption EW against mass. \citet{Du2018} measure the HIS EW to be $\sim-1.4$\AA. We find the HIS EW of our dwarf galaxies to be about 70\% as strong ($\rm EW_{HIS} =-0.99\pm0.24$\AA). At higher redshift, \citet{Jones2012} measure a comparable value ($\rm EW_{HIS}\approx -1.5$) to \citet{Du2018} at similar mass. This is consistent with the higher redshift samples of \citet{Du2018}, which suggest the HIS absorption strength does not change with redshift. \citet{Shapley2003} show that the LAEs of their sample have weaker HIS absorption than do the LBGs of their sample. It may be the case that this difference in their sample merely reflects a difference in the mass of LAEs and LBGs.

We have shown that the Ly$\alpha$ EW increases with decreasing mass in agreement with other works in the literature. As such we may expect that a sample of LAEs will have lower masses than a higher mass LBG sample. As such, the trends in $\rm EW_{HIS}$ with $\rm EW_{Ly\alpha}$ shown by \citet{Shapley2003} and \citet{Du2018} may suggest that $\rm EW_{HIS}$ is mass dependent. Both of these works do show a decrease in the absolute value of $\rm EW_{HIS}$ for their highest $\rm EW_{Ly\alpha}$ bins, though each argues that there data is consistent with no trend. To disentengle whether our decrease in $\rm EW_{HIS}$ is do to an increase in $\rm EW_{Ly\alpha}$ or mass we measure $\rm EW_{HIS}$ in our two $\rm EW_{Ly\alpha}$ sub-stacks. The low Ly$\alpha$ EW stack measures $\rm EW_{HIS}=-0.86\pm0.50$\AA\ and the high Ly$\alpha$ EW stack measures $\rm EW_{HIS}=-0.73\pm0.20$\AA\. Which are consistent with no trend in $\rm EW_{HIS}$ with $\rm EW_{Ly\alpha}$ in agreement with the statements of \citet{Shapley2003} and \citet{Du2018}. However, our measured values in our sub-samples differ from the higher mass sample of \citet{Du2018} in their figure 5. Furthermore, our $\rm EW_{HIS}$ is consistent with the highest $\rm EW_{Ly\alpha}$ bin in their samples. This points to a mass dependence, rather than a $\rm EW_{Ly\alpha}$ in our sample. Given the correlation between Ly$\alpha$ EW and mass it is likely the case that the highest $\rm EW_{Ly\alpha}$ samples of \citet{Shapley2003} and \citet{Du2018} are showing a mass dependent trend in $\rm EW_{HIS}$ as well. Together, this points to $\rm EW_{HIS}$ depending on mass, rather than the Ly$\alpha$ EW.  While the HIS EW appears to be anti-correlated with the mass of the galaxy, the HIS velocity offset is less clear. 

We measure the velocity offset of our SiIV absorption lines to be $-71\pm67\;\rm km\; s^{-1}$ and that of our CIV absorption lines to be $-449\pm60\;\rm km\; s^{-1}$. However, we caution that the velocity offset of the CIV line is complicated by possible "filling-in" of the absorption feature by emission line features closer to the systemic veloctiy. The velocity offset of the SiIV line differs significantly from that measured in the literature \citep[$-220^{+150}_{-100}$,][]{sugahara2019}. Therefore, the outflow velocities of high-ions are lower in lower-mass galaxies.

\citet{Jones2012} suggest the use of the Si IV $\rm\lambda\lambda$1393, 1402 doublet EWs as a means of tracing whether the HIS absorption is optically thick. If the ratio $\rm W_{1393}/W_{1402} \sim 2$ then the gas is optically thin, but if $\rm W_{1393}/W_{1402} \sim 1$ then the gas is optically thick. \citet{Jones2012} show a ratio of $1.4\pm0.4$ and argue that this means there is a significant amount of optically thick HIS gas present in their average galaxy. For our composite spectrum we measure $W_{1393}/W_{1402} = 1.55\pm0.76$, consistent with the SiIV gas being either optically thin or optically thick. \citet{Jones2012} measure the SiIV gas of their sample to be optically thick with low significance ($1.4\pm0.4$). Additionally, \citet{Du2018} show that their SiIV doublet is consistent with being optically thin in their redshift 2 sample ($2.13\pm0.13$). Given the low significance of the \citet{Jones2012}, the results of \citet{Du2018}, and the measurements of this paper the properties of the SiIV gas remain unclear. More data is needed to draw any conclusions regarding the opacity of the SiIV gas.

\subsection{Oligarchs vs. Dwarfs}
A favored model for the reionization of neutral hydrogen in the early universe has been the "democratic" model \citep{Naidu2020} by large quantities of high Lyman continuum escape fraction dwarf galaxies \citep[e.g.][]{Livermore2017,Endsley2023}. However, it has been suggested by \citet{Naidu2020} that it is not dwarf galaxies, but the "oligarchs" that are responsible for reionization. Their model shows that the low-mass objects contribute less to produce a rapid late reionization. However, our findings in this work and in \citet{Snapp-Kolas2022} have shown that dwarf galaxies will on average have larger Ly$\alpha$ escape fractions and lower neutral hydrogen covering fractions (which implies larger Lyman continuum escape). While the models of \citet{Naidu2020} focus on higher redshift galaxies our findings in this work in comparison with \citet{Du2018} imply that galaxies with masses at and below $10^8\;\rm M_\odot$ at higher redshift will have greater Ly$\alpha$ escape fractions and likely greater Lyman continuum escape fractions. These results appear to support the democratic model for reionization. 
\section{Summary}
\label{Summary}
In this paper we have studied the UV spectroscopic properties of typical dwarf galaxy at $z\sim 2$ via stacking. The systemic redshift of the individual galaxies is measured from our rest-optical MOSFIRE spectra. We fit the Ly$\alpha$ emission line and the LIS and HIS absorption lines using MCMC routines to measure the velocity offsets of these lines relative to the systematic redshift. We measure the EW of Ly$\alpha$ using the methods of \citet{Du2018}, and we measure the EWs of the LIS and HIS absorption lines using the best fit model of the MCMC routines. We find the following primary results from these measurements:
\begin{itemize}
    \item We find that the typical $\rm EW_{Ly\alpha}$ is much larger for dwarf galaxies than for the higher mass galaxy sample of \citet{Du2018}. This is in agreement with our earlier work \citep{Snapp-Kolas2022}. 
    \item Lower mass galaxies have about 60\% of the $\rm EW_{LIS}$ of more massive galaxies. We fit an anti-correlation and find the following best fit values for the linear model: ${\rm EW_{LIS}} = (-0.39\pm0.12){\rm log(M^*/M_\odot)} + (2.12\pm1.20)$.
    \item We find that lower-mass galaxies have lower$\rm EW_{LIS}$ at a fixed $\rm EW_{Ly\alpha}$ than high-mass galaxies. This may be connected to dwarf galaxies having lower metallicities \citep{Du2021,gburek2023}.
    \item We find the LIS gas to be optically thick ($\rm EW_{SiII1260}/EW_{SiII1526} = 1.0\pm0.3$) which implies that our measurements of the $\rm EW_{LIS}$ are still tracing the covering fraction.
    \item We find velocity offsets in the LIS absorption lines for lower mass galaxies to be lower than in high mass galaxies in agreement with the FIRE simulations \citep{Muratov2015}. 
    \item The $\rm EW_{HIS} = -0.99\pm0.24$\AA\ is smaller in absolute value than for higher mass galaxies. We find this to be true even in our $\rm EW_{Ly\alpha}$ sub-stacks, making it clear that this is a mass dependent effect and not do to the galaxies possessing higher $\rm EW_{Ly\alpha}$.
\end{itemize}

It is clear that dwarf galaxies have higher Ly$\alpha$ EWs, weaker LIS and HIS absorption EWs, and lower LIS velocity offsets relative to more massive galaxies. Larger Ly$\alpha$ EW's imply higher Ly$\alpha$ escape fractions \citep[e.g.][]{Yang2017e} which may imply larger Lyman continuum escape fractions \citep{Dijkstra2016,Verhamme2017,Izotov2020,Naidu2021,Flury2022}. Similarly, \citet{Trainor2019}, \citet{saldana-lopez2022}, and \citet{Mainali2022} show that the Ly$\alpha$ escape fraction is anti-correlated with $\rm EW_{LIS}$ suggesting higher Lyman continuum escape for weaker $\rm EW_{LIS}$. Our dwarf galaxies exhibit stronger Ly$\alpha$ and weaker LIS EWs relative to more massive samples. Additionally, our galaxies have optically thick LIS absorption lines which means that the $\rm EW_{LIS}$ is still tracing the covering fraction and therefore there is likely greater Lyman continuum escape fractions. The cumulative evidence suggests that dwarf galaxies likely play an important role in providing the ionizing background in the early universe. 

\section*{Acknowledgments}
This research made use of {\ttfamily{PypeIt}\footnote{\url{https://pypeit.readthedocs.io/en/latest/}}},
a Python package for semi-automated reduction of astronomical slit-based spectroscopy
\citep{pypeit:joss_pub, pypeit:zenodo}. \\
Some of the data presented herein were obtained at the W. M. Keck Observatory, which is operated as a scientific partnership among the California Institute of Technology, the University of California and the National Aeronautics and Space Administration. The Observatory was made possible by the generous financial support of the W. M. Keck Foundation. \\
The authors wish to recognize and acknowledge the very significant cultural role and reverence that the summit of Maunakea has always had within the indigenous Hawaiian community.  We are most fortunate to have the opportunity to conduct observations from this mountain
Based on observations made with the NASA/ESA \textit{Hubble Space Telescope}, obtained from the Data Archive at the Space Telescope Science Institute, which is operated by the Association of Universities for Research in Astronomy, Inc., under NASA contract NAS5-26555. These observations are associated with programs \#9289, \#11710, \#11802, \#12201, \#12931, \#13389, \#14209.

\section*{Data Availability}

This paper is based on public data from the Hubble Space Telescope as well as from programs 12201, 12931, 13389, 14209.  Spectroscopic data from our survey with the Keck Observatory. These data are available upon request from Dr. Christopher Snapp-Kolas or Dr. Brian Siana.



\bibliographystyle{mnras}
\bibliography{Paper1_updated} 

\end{document}